\documentclass[12pt]{article}
\usepackage{epsfig}
\usepackage{latexsym} 
\usepackage{amssymb}  
\usepackage{amsfonts} 

\newcommand{\Z}{\mathbb{Z}}

\makeatletter
\@addtoreset{equation}{section}
\makeatother

\title{Solution of the Complex Action Problem in the Potts Model 
for Dense QCD
\footnote{This work is supported in part by funds provided by the U.S.
Department of Energy (D.O.E.) under cooperative research agreements
DE-FC02-94ER40818 and DE-FG02-96ER40945.}}
\author{M. Alford$^{{\rm a}}$, S. Chandrasekharan$^{{\rm b}}$, 
J. Cox$^{{\rm c}}$ and U.-J. Wiese$^{{\rm c}}$ \\[1.5ex]
$^{{\rm a}}$ 
  Department of Physics and Astronomy \\
  Glasgow  University \\
  Glasgow G12 8QQ, UK \\[1.5ex]
$^{{\rm b}}$ 
  Department of Physics \\
  Duke University \\
  Durham, North Carolina 27708, U.S.A. \\[1.5ex]
$^{{\rm c}}$ 
  Center for Theoretical Physics, \\
  Massachusetts Institute of Technology (MIT) \\
  Cambridge, Massachusetts 02139, U.S.A.
}
\date{} 
\begin{document}

\maketitle

\vspace{-1.0cm}

\begin{abstract} \normalsize

Monte Carlo simulations of lattice QCD at non-zero baryon chemical potential
$\mu$ suffer from the notorious complex action problem. We consider QCD with 
static quarks coupled to a large chemical potential. This leaves us with an 
$SU(3)$ Yang-Mills theory with a complex action containing the Polyakov loop. 
Close to the deconfinement phase transition the qualitative features of this 
theory, in particular its $\Z(3)$ symmetry properties, are captured by the 3-d 
3-state Potts model. We solve the complex action problem in the Potts model by 
using a cluster algorithm. The improved estimator for the $\mu$-dependent part
of the Boltzmann factor is real and positive and is used for importance 
sampling. We localize the critical endpoint of the first order deconfinement 
phase transition line and find consistency with universal 3-d Ising behavior. 
We also calculate the static quark-quark, quark-anti-quark, and 
anti-quark-anti-quark potentials which show screening as expected for a system 
with non-zero baryon density.

\end{abstract}
 
\newpage

\section{Introduction and Summary}

Non-perturbative dense QCD can presently not be studied from first principles 
because Monte Carlo simulations of lattice QCD with non-zero baryon chemical
potential $\mu$ suffer from a severe complex action problem. The Boltzmann 
factor in the path integral can then not be interpreted as a probability and 
standard importance sampling methods fail. In particular, when the 
$\mu$-dependent part of the Boltzmann factor is included in the measured 
observables, due to severe cancellations the required statistics is 
exponentially large in the space-time volume \cite{Bar98,Alf99}.

The complex action problem prevents the numerical simulation of a large class 
of interesting physical systems including other field theories at non-zero 
chemical potential or non-zero $\theta$-vacuum angle as well as some fermionic
field theories with an odd number of flavors. 
A special case of the complex action problem is
the so-called fermion sign problem which arises for fermionic path 
integrals formulated in a Fock state basis. The problem is due to paths that 
correspond to an odd permutation of fermion positions which contribute 
negatively to the path integral. 
There are numerous condensed matter systems ranging from the repulsive
Hubbard model away from half-filling to antiferromagnetic quantum spin
systems in an external magnetic field that cannot be simulated with
standard Monte Carlo algorithms. Meron-cluster algorithms have been
used to solve the sign or complex action problems in several of these
cases. For example, the first meron-cluster algorithm has led to a
solution of the complex action problem in the 2-d $O(3)$ symmetric
field theory at non-zero $\theta$-vacuum angle \cite{Bie95}. In this
model, some of the clusters are half-instantons, so they are called
meron-clusters. The complex action problem also arises in the 2-d
$O(3)$ model at non-zero chemical potential. When formulated as a
D-theory \cite{Cha97,Bea98,Wie99,Bro99} --- i.e.~in terms of discrete
variables that undergo dimensional reduction --- the complex action
problem has also been solved with a meron-cluster algorithm
\cite{Cox00a}. Recently, the meron concept has been generalized to
fermions \cite{Cha99a}. Meron-cluster algorithms have led to a
complete solution of the fermion sign problem in a variety of models
including non-relativistic spinless fermions \cite{Cha99a,newref1},
relativistic staggered fermions \cite{Cha00,Cox00b,newref2} and some models in
the Hubbard model family \cite{Cox00a,newref3}.\footnote{The models
investigated so far only show s-wave superconductivity.}
Recently, a meron-cluster algorithm has been used to solve the sign
problem that arises for quantum antiferromagnets in an external
magnetic field \cite{Cha99b}. For a review of these recent
developments and a preliminary version of the present results 
see \cite{newref4,newref5}.

In the conventional formulation of lattice QCD the quarks are represented by Grassmann 
fields. When the quarks are integrated out, they leave behind a fermion 
determinant that acts as a non-local effective action for the gluons. At zero 
chemical potential and for an even number of flavors, the fermion determinant 
is real and positive and can thus be interpreted as a probability for 
generating gluon field configurations. Despite the fact that standard 
importance sampling techniques apply, the non-local nature of the effective 
gluon action makes lattice QCD simulations with dynamical fermions very 
time consuming. With a non-zero chemical potential for the baryon number, the 
fermion determinant becomes complex and standard importance sampling techniques
fail completely \cite{Bar98,Alf99}. This is the reason why non-perturbative QCD
at non-zero baryon density can presently not be studied from first principles.

It is natural to ask if a meron-cluster algorithm could be used to solve the 
complex action problem in QCD. 
When one integrates out light quarks, one obtains a
non-local effective action for the gluons and it appears
unlikely that the meron concept will apply.
On the other hand, when one describes the 
quarks in a Fock state basis, the complex action problem is still
present, in the form of a fermion sign problem. 
Our hope is that this problem will
eventually be solved by a meron-cluster algorithm applied to the
D-theory formulation of QCD \cite{Cha97,Bea98,Wie99,Bro99}, since
the quark and gluon degrees of freedom are then
discrete and are much
easier to handle.
In this
paper we address a simpler situation first. We consider QCD in the limit
of very heavy quarks with a large chemical potential. 
These can be integrated out, introducing Polyakov loops
into the effective gluon action.
When quarks are integrated out at non-zero chemical potential $\mu$
we expect a complex action, and in this case it arises
because a Polyakov loop $\Phi$ and its charge conjugate $\Phi^*$ get
different weights when $\mu\neq 0$.

Polyakov loops are only non-local in the Euclidean time direction,
so this effective gluon action is
more manageable than the one that arises for a general fermion
determinant. Indeed, Blum, Hetrick and Toussaint have simulated the
theory in this form on lattices of moderate size where the complex
action problem is less severe \cite{Blu96}. Recently, Engels,
Kaczmarek, Karsch and Laermann have studied QCD with heavy quarks at
fixed baryon number. Again, for moderate baryon density and moderate
volumes the complex action problem is not too severe and simulations
are possible \cite{Eng99}. 
Ultimately one would like to be able to solve 
the complex action problem for this gluon action completely. At the moment,
we still cannot apply a meron-cluster algorithm to solve the problem, 
because the construction of efficient cluster algorithms for non-Abelian 
gauge theories seems to be impossible for Wilson's formulation of lattice 
field theory. Here  we will simplify the problem further
by replacing the gauge dynamics by that of the $\Z(3)$ Potts model 
representing the Polyakov loops \cite{HaKa83,DeDe83}. 
We have found a cluster algorithm that 
solves this complex action problem in the Potts model approximation to QCD.

The 3-d $\Z(3)$-symmetric Potts model has often been used as an
approximation to QCD with static quarks. In particular, the phase
transition to a broken $\Z(3)$ symmetry phase at high temperature
corresponds to the first order deconfining phase transition in QCD. As
has been noted by Condella and DeTar, a term that corresponds to a
chemical potential can also be included in the Potts model, explicitly
breaking the $\Z(3)$ symmetry \cite{Con00}. 
As the coefficient of this term grows,
the first order deconfinement phase transition persists but it becomes
weaker and ultimately disappears in a critical endpoint. This point is
expected to be in the universality class of the 3-d Ising model. 
In this paper we will confirm this expectation with numerical
simulations.

In principle, one can imagine deriving an effective 3-d 3-state Potts
model directly from QCD by integrating out all degrees of freedom
except for the $\Z(3)$ phase of the Polyakov loop.  However, the
resulting Potts model action would be very complicated and cannot be
derived in practice, except in the strong coupling limit. Here we
approximate QCD with heavy quarks by a 3-d $\Z(3)$-symmetric Potts
model with a standard nearest-neighbor interaction. Universal features
like the nature of the critical endpoint of the deconfinement phase
transition are correctly reproduced in this approximation. Figure 1
contains the phase diagram of the
\begin{figure}
\begin{center}
\epsfig{file=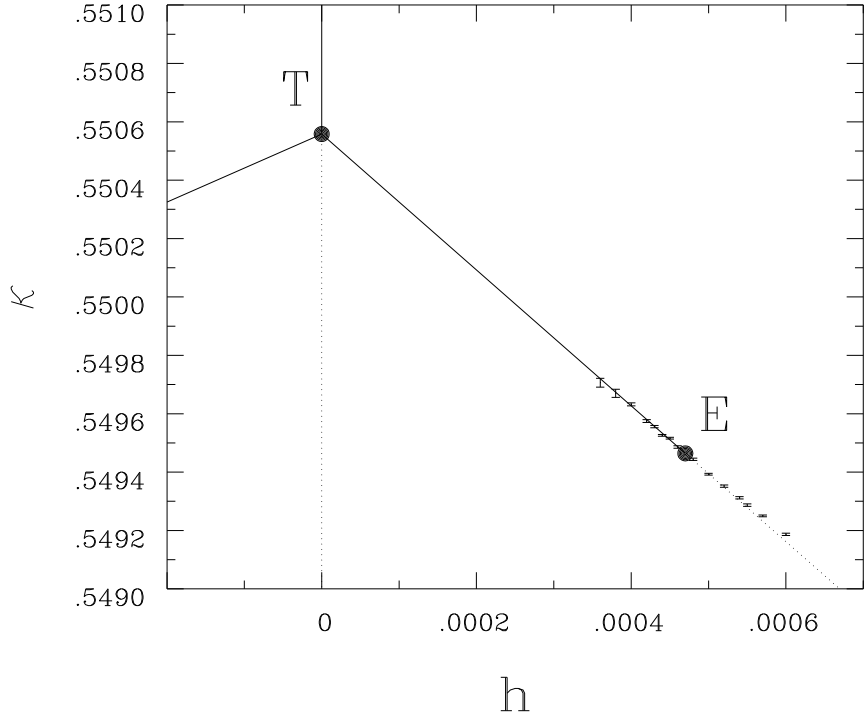,
width=9.5cm,angle=90,
bbllx=50,bblly=200,bburx=535,bbury=800}
\end{center} 
\vspace*{-0.6cm}
\caption{\it The phase diagram of the $\Z(3)$ Potts model
in the $(h,\kappa)$-plane. The ordinary 
deconfinement phase transition at $T=(0,0.550565(10))$ is a triple point  from which a
line of first order phase transitions emerges. This line terminates in the 
critical endpoint $E=(0.000470(2),0.549463(13))$ 
and continues only as a crossover.}
\end{figure}
3-d 3-state Potts model in the $(h,\kappa)$-plane. 
The parameter $h$ represents $\exp(\beta(\mu-M))$ in QCD with quarks
of mass $M$ at chemical potential $\mu$.  We study the limit $M,\mu
\to \infty$ for any given $\mu - M$. Large $h$ corresponds to $\mu> M$
and small $h$ to $\mu < M$.  Because $\mu-M \ll M,\mu$ we are always,
for any $h$, in the immediate neighborhood of the onset of non-zero
density for the heavy quarks.  This means that it does not matter
whether they are fermions or bosons, since they never move.
The difference would only become apparent above the onset, 
where either a Fermi surface or a degenerate Bose gas would occur,
and our order of limits is such that we never get that far from the onset.
The parameter $\kappa$ is the standard Potts model coupling,
which corresponds roughly to the temperature $T=1/\beta$.
The ordinary first-order deconfinement phase transition at $h=0$ 
(point $T$ in Figure 1) extends into a line of first order
transitions that terminates in the critical endpoint $E$. 
This endpoint occurs at such a low value of $h$ that the
complex action problem is not very severe there, and
we found that the most efficient way to locate and study it
was to employ a reweighted Metropolis algorithm, which can in this case
be applied at volumes large enough to show the critical behavior.
Similar methods were used recently by Karsch and Stickan \cite{Kar00}
in a version of the 3-d 3-state Potts model where the action
is real, and the endpoint was found to 
have the critical exponents of the 3-d Ising model. 
We find that in the Potts model with complex action the endpoint has
the same critical properties. Furthermore its position is barely
shifted in comparison to the model with real action.
In this paper we do not limit our attention to the endpoint, but
develop a method that solves the complex action problem everywhere in the
phase diagram.

We also calculate the potentials between static quarks and anti-quarks
in the Potts model approximation to QCD. In the confined phase at $\mu
= 0$ the static quark-anti-quark potential is linearly rising with the
distance as a manifestation of confinement. For the same reason the
quark-quark and anti-quark-anti-quark potentials are infinite at all
distances. In the deconfined phase the quark-anti-quark potential
reaches a plateau at twice the (now finite) free energy of a
quark. Similarly, the quark-quark and anti-quark-anti-quark potentials
are no longer infinite. It should be noted that quark-quark and
anti-quark-anti-quark potentials are usually not calculated in lattice
simulations. This is because---as a consequence of the $\Z(3)$ Gauss
law---quark or anti-quark pairs cannot exist in a finite spatial
volume with periodic boundary conditions \cite{Hil83}. Interestingly,
this changes for $\mu \neq 0$ because then there are compensating
background charges in the medium that can absorb the $\Z(3)$ 
flux of an external
quark. Since the chemical potential explicitly breaks the $\Z(3)$
symmetry, there is no longer a clear distinction between confinement
and deconfinement for $\mu \neq 0$. This manifests itself in the phase
diagram by the fact that confined and deconfined phases are
analytically connected. 
\begin{figure}
\begin{center}
\epsfig{file=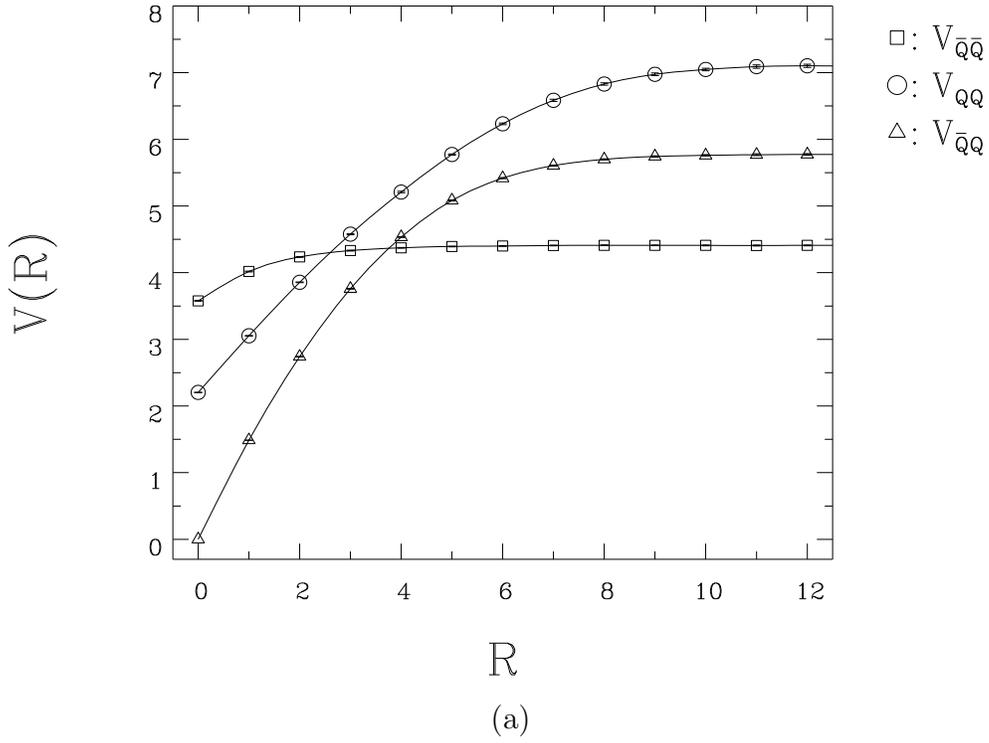,
width=9.0cm,angle=90,
bbllx=50,bblly=200,bburx=535,bbury=670}\\[1ex]
(a)\\[3ex]
\epsfig{file=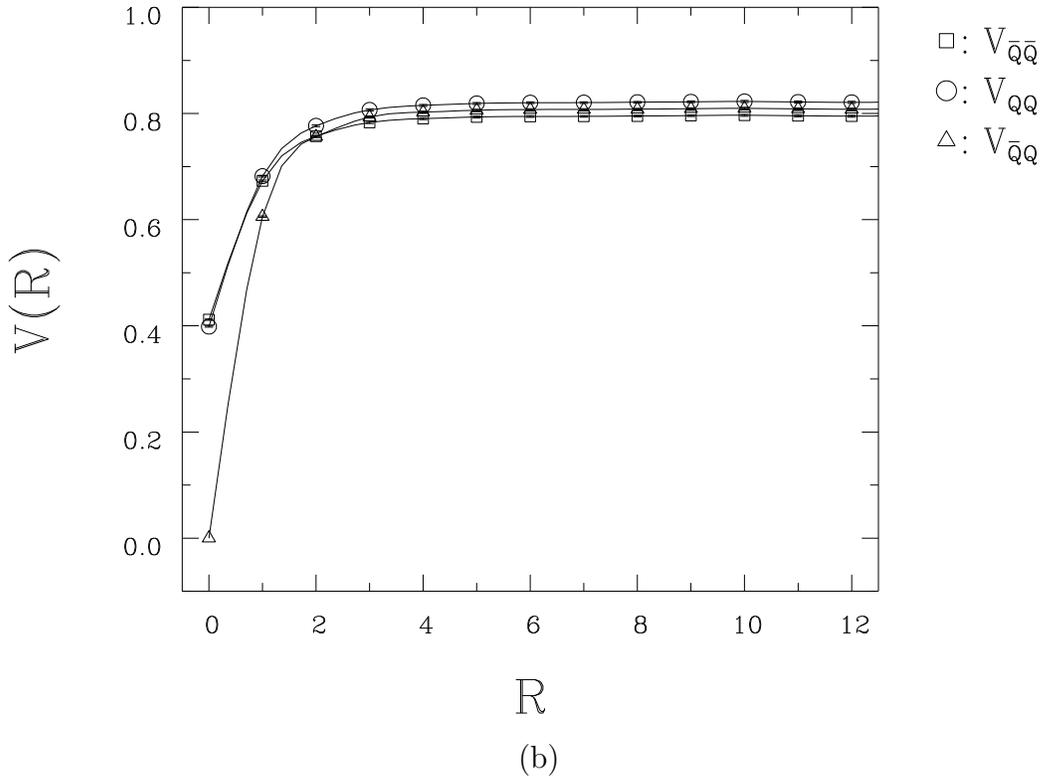,
width=9.5cm,angle=90,
bbllx=50,bblly=200,bburx=535,bbury=670}\\[1ex]
(b)
\end{center} 
\vspace*{-0.6cm}
\caption{\it The static quark-anti-quark, quark-quark and anti-quark-anti-quark
potentials (a) on the confined side (at $h = 0.01$, $\kappa = 0.50$) and (b) on 
the deconfined side (at $h = 0.01$ and $\kappa = 0.56$) of the crossover.}
\end{figure}
Figure 2 shows the quark-anti-quark, quark-quark and anti-quark-anti-quark 
potentials on the confined side (a) and on the deconfined side (b) of the 
crossover. Note that at $\mu \neq 0$ even in the confined phase the 
quark-anti-quark potential now reaches a plateau. 
The plateau height 
corresponds to the sum of the free energies $F_Q$ of an external quark 
and $F_{\bar Q}$ 
of an external anti-quark. For $\mu > 0$ quarks are favored in the medium
while anti-quarks are 
suppressed. As a consequence, the free energy of an
external static quark is 
larger than that of an external static anti-quark. While an external static 
anti-quark can bind with a 
single background quark from the medium and form a meson, an external static 
quark needs two quarks from the medium to form a baryon. Indeed, on the 
confined side of the transition $F_Q$ is clearly larger than $F_{\bar Q}$, 
while on the deconfined side $F_Q$ and $F_{\bar Q}$ are more or less the same.
We have normalized the potentials such that at zero distance a static 
quark-anti-quark pair has zero energy. In the Potts model, two quarks at zero 
distance are indistinguishable from a single anti-quark, and similarly, two 
anti-quarks on top of each other behave like a single quark. Hence, at zero 
distance the quark-quark potential $V_{Q Q}(0)$ agrees with the free energy of 
a single anti-quark $F_{\bar Q}$ and the anti-quark-anti-quark potential obeys 
$V_{\bar Q \bar Q}(0) = F_Q$. At asymptotic distances the potentials 
$V_{Q \bar Q}(\infty)$, $V_{Q Q}(\infty)$ and $V_{\bar Q \bar Q}(\infty)$ 
take the values $F_Q + F_{\bar Q}$, $2 F_Q$ and $2 F_{\bar Q}$, respectively. 
This behavior is consistent with our numerical data shown in fig.2.

In the absence of a chemical potential, the Potts model can be simulated with 
the original Swendsen-Wang cluster algorithm \cite{Swe87}. When a chemical 
potential is introduced, the Potts model 
suffers from the complex action problem and standard importance
sampling methods including the cluster algorithm fail. In this paper,
we construct an improved estimator for the $\mu$-dependent part of the
Boltzmann factor by averaging analytically over all configurations
related to each other by cluster flips. In contrast to the original
Boltzmann factor, the improved estimator is real and positive and can
be used for importance sampling. This solves the complex action
problem completely.

Although the Potts model inherits the complex action problem from QCD, it can 
be transformed into a ``flux model'' that has no complex action problem 
\cite{Con00}. The flux model has been simulated in \cite{Con00} and the 
disappearance of the first order deconfining phase transition at large chemical
potential has been observed numerically. These results may give encouragement 
to the hope that QCD itself could be transformed into a model without a complex
action problem. 
In this paper we show that, at least for the Potts model, it is
more efficient to leave it in its usual form and solve the 
complex action problem with our cluster algorithm than to transform
it into a flux model and use conventional Metropolis methods.

The paper is organized as follows. Section 2 contains a derivation of the
effective gluon action resulting from static quarks with large chemical 
potential as well as its Potts model approximation. In section 3 we describe 
the cluster algorithm that solves the complex action problem. Section 4 
contains the derivation of the flux representation of the Potts model and a 
description of a Metropolis algorithm to simulate it. A comparison of the 
Metropolis algorithm for the flux model and the cluster algorithm for the
original Potts model shows that the latter is more efficient. In section 5 we 
present the physical results concerning the critical endpoint $E$.
Using 
finite-size scaling techniques, we are able to determine the position of the 
critical endpoint of the deconfinement phase transition to high accuracy. Our 
results are consistent with the expected universal 3-d Ising behavior. Finally,
section 6 contains our conclusions.

\section{QCD with Heavy Quarks and the 3-d 3-State Potts Model}

The partition function for a pure $SU(3)$ Yang-Mills theory is given by
\begin{equation}
Z = \int {\cal D}A \ \exp(- S[A]),
\end{equation}
where
\begin{equation}
S[A] = \int_0^\beta dt \int d^3x \ \frac{1}{2 g^2} \mbox{Tr}
[F_{\mu\nu} F_{\mu\nu}],
\end{equation}
is the Euclidean action for the gluons and $\beta$ is the inverse temperature.
The action is invariant under gauge transformations
\begin{equation}
\label{gauge}
g(\vec x,0) = g(\vec x,\beta) z, 
\end{equation}
that are periodic in Euclidean time up to an element $z$ of the center 
$\Z(3) = \{\exp(2 \pi i n/3), n = 1,2,3\}$ of the non-Abelian gauge group. In 
the presence of a single external heavy quark of bare mass $M$ at an 
undetermined position $\vec x$ the partition function turns into
\begin{equation}
Z_Q = \int {\cal D}A \ \Phi[A] \exp(- S[A]) \exp(- \beta M),
\end{equation}
where
\begin{equation}
\Phi[A] = \int d^3x \ \mbox{Tr}[{\cal P} \exp(- \int_0^\beta dt \ 
A_4(\vec x,t))],
\end{equation}
is the spatial integral of the Polyakov loop. Ultimately, the mass $M$ will be 
sent to infinity. Note that while the center transformation of eq.(\ref{gauge})
leaves the pure gluon action $S[^g\!A] = S[A]$ invariant, the Polyakov loop 
transforms into
\begin{equation}
\Phi[^g\!A] = z \Phi[A].
\end{equation}
This shows that in the presence of the external quark, the $\Z(3)$ symmetry is 
explicitly broken. The partition function for a system of gluons in the 
presence of a single heavy anti-quark is given by
\begin{equation}
Z_{\bar Q} = \int {\cal D}A \ \Phi[A]^* \exp(- S[A]) \exp(- \beta M),
\end{equation}
where $*$ denotes complex conjugation. Let us now consider a system of gluons 
in a background of $n$ static quarks and $\bar n$ static anti-quarks. The 
partition function then takes the form
\begin{equation}
Z_{n,\bar n} = \int {\cal D}A \ \frac{1}{n!} \Phi[A]^n 
\frac{1}{\bar n !} (\Phi[A]^*)^{\bar n} \exp(- S[A]) 
\exp(- \beta M(n + \bar n)).
\end{equation}
The factors $1/n!$ and $1/\bar n !$ appear because 
quarks are indistinguishable, as are antiquarks.
Introducing the quark chemical 
potential $\mu$ that couples to $(n - \bar n)$, i.e. three times
the baryon number,
we obtain the grand canonical partition function
\begin{eqnarray}
Z(\mu)&=&\sum_{n,\bar n} Z_{n,\bar n} \exp(\beta \mu (n - \bar n)) 
\nonumber \\
&=&\sum_{n,\bar n} \int {\cal D}A \ \frac{1}{n!} \Phi[A]^n 
\frac{1}{\bar n !} (\Phi[A]^*)^{\bar n} \exp(- S[A]
- \beta n (M - \mu) - \beta \bar n (M + \mu)) \nonumber \\ 
&=&\int {\cal D}A \ \exp(- S[A] + \exp(- \beta (M - \mu)) \Phi[A] +
\exp(- \beta (M + \mu)) \Phi[A]^*). \nonumber \\ \
\end{eqnarray}
As expected, the presence of quarks and anti-quarks leads to an explicit
breaking of the $\Z(3)$ center symmetry. Furthermore, in the presence of a
non-zero chemical potential the effective action for the gluons is complex. 
Note that in the $SU(2)$ case the action remains real because then the Polyakov
loop itself is real, i.e.~$\Phi[A]^* = \Phi[A]$. The action becomes real even
in the $SU(3)$ case if $\mu$ is purely imaginary. Furthermore, one can see that
the chemical potential explicitly breaks the charge conjugation symmetry that 
replaces $\Phi[A]$ by $\Phi[A]^*$. In fact, under charge conjugation the action
turns into its complex conjugate. We have assumed that the quarks are static. 
Hence, to be consistent we must consider the limit $M \rightarrow \infty$. In 
order to obtain a non-trivial result, we simultaneously take the limit 
$\mu \rightarrow \infty$ such that $M - \mu$ remains finite. The partition 
function then simplifies to
\begin{equation}
Z(\mu) = \int {\cal D}A \ \exp(- S[A] + \exp(- \beta (M - \mu)) \Phi[A]).
\end{equation}
As discussed in \cite{Blu96} and \cite{Eng99}, a similar result can be obtained
by simplifying the full QCD quark determinant in the static quark limit. In 
general the determinant would contain all Wilson loops, but because $M$ is 
large most of them are suppressed. The only ones that survive are those for 
which the enhancement due to the chemical potential compensates for the 
suppression due to the mass. These are the Polyakov loops that progress in a 
straight line from Euclidean time $t = 0$ to $t = \beta$ at some position 
$\vec x$. In the loop expansion of the quark determinant, each of these has a 
weight $\exp(- \beta(M - \mu))$. 

Up to this point we have treated QCD consistently in the static quark limit.
The resulting effective action for the gluons is complex and we presently don't
know how to simulate it efficiently. For that reason we now replace the gluon 
system by a simple 3-d lattice 3-state Potts model. The Potts spins 
$\Phi_x \in \Z(3)$ replace the original Polyakov loop variables and the 
partition function turns into
\begin{equation}
\label{Potts}
Z(h) = \int {\cal D}\Phi \ \exp(- S[\Phi] + h \sum_x \Phi_x),
\end{equation}
where $h$ replaces $\exp(- \beta(M - \mu))$. Note that the Potts model action 
is still complex. In principle, one can imagine integrating out all QCD 
degrees of freedom except for the $\Z(3)$ phase of the Polyakov loop and thus 
derive an effective Potts model action directly from QCD. In practice this is 
impossible, except in the strong coupling limit. For simplicity, we therefore 
replace the pure gluon action $S[A]$ by a standard nearest-neighbor Potts model
interaction
\begin{equation}
S[\Phi] = - \kappa \sum_{x,i} \delta_{\Phi_x,\Phi_{x+\hat i}}.
\end{equation}
The coupling constant $\kappa$ is not 
related in a simple way to the parameters of 
QCD. Still, a large value of $\kappa$ corresponds qualitatively to the 
high-temperature deconfined phase, while small $\kappa$ values correspond to 
the confined phase. As mentioned in the introduction, the Potts model also 
retains the general features of the QCD phase diagram. At $h = 0$ ($M$ 
infinite, $\mu$ finite) there is a first-order phase transition as a function 
of $\kappa$, between the disordered (confined) phase that respects the $\Z(3)$ 
symmetry and the ordered (deconfined) phase that spontaneously breaks it. An 
order parameter for this transition is $\langle \Phi \rangle$. As $h$ rises 
from zero, the chemical potential term explicitly breaks the $\Z(3)$ symmetry, 
the phase transition weakens, and then ends at a critical point. 
Correspondingly, in heavy-quark QCD the quarks begin to contribute to the 
partition function when $\mu$ gets close to $M$, and there is no longer an 
order parameter for deconfinement. The deconfining phase transition terminates 
at a critical endpoint.

\section{Cluster Algorithm Solution of the Complex Action Problem}

In this section we first discuss the general nature of the complex action 
problem and then discuss the cluster algorithm that solves this problem for the
Potts model. We also construct improved estimators for various physical
quantities.

\subsection{The General Nature of the Complex Action Problem}
 
When the action is complex the resulting Boltzmann factor cannot be interpreted
as a probability and hence standard importance sampling techniques fail. When 
one uses just the absolute value of the Boltzmann factor for importance 
sampling and includes its complex phase in measured observables $O$, 
expectation values take the form
\begin{eqnarray}
\label{observable}
\langle O \rangle&=&\frac{1}{Z} \int {\cal D}\Phi \ O[\Phi] 
\exp(- S[\Phi] + h \sum_x \Phi_x) \nonumber \\
&=&\frac{\langle O \exp(i h \sum_x \mbox{Im}\Phi_x) \rangle_R}
{\langle \exp(i h \sum_x \mbox{Im}\Phi_x) \rangle_R}.
\end{eqnarray}
The subscript $R$ refers to a modified ensemble with a real action described by
the partition function
\begin{equation}
Z_R = \int {\cal D}\Phi \ \exp(- S[\Phi] + h \sum_x \mbox{Re}\Phi).
\end{equation}
By definition we have
\begin{eqnarray}
\label{phase}
\langle \exp(i h \sum_x \mbox{Im}\Phi_x) \rangle_R&=&\frac{1}{Z_R} 
\int {\cal D}\Phi \ \exp(i h \sum_x \mbox{Im}\Phi_x) 
\exp(- S[\Phi] + h \sum_x \mbox{Re}\Phi) \nonumber \\
&=&\frac{Z}{Z_R} \approx \exp(- V (f - f_R)),
\end{eqnarray}
where $f$ and $f_R$ are the free energy densities of the original complex and 
the modified real action systems, respectively, and $V$ is the spatial volume. 
Hence, the denominator in eq.(\ref{observable}) becomes exponentially small as 
one increases the volume. The same is true for the numerator, because 
$\langle O \rangle$ itself is not exponentially large in $V$.

Although, in principle, simulating the modified ensemble is correct, in 
practice this method fails for large volumes. The reason is that observables 
are obtained as ratios of exponentially small numerators and denominators which
are themselves averages of quantities of order one. This leads to very severe 
cancellations and requires an exponentially large statistics in order to obtain
accurate results. To see this, we estimate the relative statistical error in 
the determination of the average phase of the Boltzmann factor 
$\exp(i h \sum_x \mbox{Im}\Phi_x)$. Since 
$\langle \exp(i h \sum_x \mbox{Im} \Phi_x) \rangle_R = Z/Z_R$ the average 
itself is real. When one generates $N$ statistically independent field 
configurations in a Monte Carlo simulation, the resulting error to signal ratio
is given by
\begin{eqnarray}
\frac{\Delta \exp(i h \sum_x \mbox{Im}\Phi_x)}
{\langle \exp(i h \sum_x \mbox{Im}\Phi_x) \rangle_R}&=&
\frac{\sqrt{\langle |\exp(i h \sum_x \mbox{Im}\Phi_x) - 
\langle \exp(i h \sum_x \mbox{Im}\Phi_x) \rangle_R|^2 \rangle_R}}{\sqrt{N} 
\langle \exp(i h \sum_x \mbox{Im}\Phi_x) \rangle_R} \nonumber \\
&=&\frac{\sqrt{1 - \langle \exp(i h \sum_x \mbox{Im}\Phi_x) \rangle_R^2}}
{\sqrt{N} \langle \exp(i h \sum_x \mbox{Im}\Phi_x) \rangle_R} 
\approx\frac{\exp(V (f - f_R))}{\sqrt{N}}.
\end{eqnarray}
For large $V$ we have used $\langle \exp(i h \sum_x \mbox{Im}\Phi_x) \rangle_R
\ll 1$ as implied by eq.(\ref{phase}). Consequently, in order to obtain an 
acceptable error to signal ratio one must generate at least 
$N \approx \exp(2 V (f - f_R))$ configurations. For large volumes this is
impossible in practice. 

\subsection{The Cluster Algorithm for the Potts Model}

Let us now outline the ideas that underlie the cluster algorithm that we use to
solve the complex action problem. It is based on the original Swendsen-Wang
cluster algorithm \cite{Swe87} for the Potts model without chemical potential.
In fact, in the limit $h = 0$ our algorithm reduces to that algorithm. The 
Swendsen-Wang cluster algorithm decomposes the lattice into independent
clusters of connected sites. Each spin belongs to exactly one cluster and all
spins within a cluster are assigned the same random $\Z(3)$ element. In this 
paper, we construct an improved estimator for the $h$-dependent part 
$\exp(h \sum_x \Phi_x)$ of the Boltzmann factor by analytically averaging it
over all configurations related to each other by cluster flips. Although, for 
an individual configuration $\exp(h \sum_x \Phi_x)$ is in general complex, its 
improved estimator is always real and positive and can thus be used for 
importance sampling. This completely solves the complex action problem.

Let us first describe the original Swendsen-Wang algorithm for $h = 0$. In this
method one introduces variables $b = 0,1$ for each bond connecting neighboring 
lattice sites $x$ and $y = x + \hat i$ and one writes the nearest neighbor 
Boltzmann factor as
\begin{equation}
\exp(\kappa \delta_{\Phi_x,\Phi_y}) = \sum_{b = 0,1}
[\delta_{b,1} \delta_{\Phi_x,\Phi_y} (e^\kappa - 1) + \delta_{b,0}].
\end{equation}
In the enlarged configuration space of spin and bond variables, the bond 
variables impose constraints between the spin variables. When a bond is put 
(i.e.~when $b = 1$), the spin Boltzmann factor is $\delta_{\Phi_x,\Phi_y} 
(e^\kappa - 1)$ and hence the spin variables $\Phi_x$ and $\Phi_y$ at the two 
ends of the bond must be identical. On the other hand, when the bond is not put
($b = 0$), the spin Boltzmann factor is $1$ and thus the variables $\Phi_x$ and
$\Phi_y$ fluctuate independently. The spin variables, in turn, determine the 
probability to put a bond. When the spins $\Phi_x$ and $\Phi_y$ are different, 
the bond Boltzmann factor is $\delta_{b,0}$ and thus the bond is not put. On 
the other hand, when $\Phi_x$ and $\Phi_y$ are the same, the bond Boltzmann 
factor is $[\delta_{b,1} (e^\kappa - 1) + \delta_{b,0}]$. Consequently, a bond 
between parallel spins is put with probability $p = 1 - e^{- \kappa}$. Note 
that for $\kappa = 0$ no bonds are put, while for $\kappa = \infty$ parallel 
spins are always connected by a bond.

The Swendsen-Wang cluster algorithm updates bond and spin variables in 
alternating order. First, for a given spin configuration, bonds are put with 
probability $p$ between parallel neighboring spins. No bonds are put between 
non-parallel spins. Then the spins are updated according to the constraints 
represented by the resulting bond configuration. Spins connected by bonds must 
remain parallel, while spins not connected by bonds fluctuate independently. 
Hence, to update the spins, one must identify clusters, i.e.~sets of spins that
are connected by bonds. All spins in a cluster are parallel and are assigned 
the same random $\Z(3)$ element in the spin update. All spins belong to exactly
one cluster. It should be noted that a cluster may consist of a single spin. A 
configuration consisting of $N_C$ clusters can be viewed as a member of a 
sub-ensemble of $3^{N_C}$ equally probable configurations which result by
assigning $\Z(3)$ elements to the various clusters in all possible ways. As was
already pointed out by Swendsen and Wang, one can construct improved estimators
for various physical quantities by averaging analytically over all $3^{N_C}$
configurations in a sub-ensemble. Since the number of clusters is proportional
to the volume, this effectively increases the statistics by a factor that is
exponentially large in $V$.

Let us construct an improved estimator for the $h$-dependent part 
$\exp(h \sum_x \Phi_x)$ of the Boltzmann factor. Although for an individual 
configuration this term is in general complex, its average over a sub-ensemble 
of $3^{N_C}$ configurations is always real and positive. This results from the 
following observations. The $h$-dependent part of the Boltzmann factor is a 
product of cluster contributions
\begin{equation}
\exp(h \sum_x \Phi_x) = \prod_C \exp(h \sum_{x \in C} \Phi_x).
\end{equation}
Since the clusters are independent, the sub-ensemble average is a product
\begin{equation}
\langle \exp(h \sum_x \Phi_x) \rangle_{3^{N_C}} = 
\prod_C \langle \exp(h \sum_{x \in C} \Phi_x) \rangle_3,
\end{equation}
of 3-state averages for the individual clusters
\begin{eqnarray}
\label{average}
\langle \exp(h \sum_{x \in C} \Phi_x) \rangle_3&=& 
\frac{1}{3} \sum_{\Phi \in \Z(3)} \exp(h |C| \Phi) \nonumber \\
&=&\frac{1}{3} [\exp(h |C|) + 2 \exp(- h |C|/2) \cos(\sqrt{3} h |C|/2)]
\nonumber \\
&=&W(C),
\end{eqnarray}
which defines a weight $W(C)$ for each cluster. We have used the fact that all 
spins $\Phi_x$ in a given cluster $C$ take the same value $\Phi \in \Z(3)$ so 
that $\sum_{x \in C} \Phi_x = |C| \Phi$ where $|C| = \sum_{x \in C} 1$ is the 
cluster size. It is easy to show that the expression in eq.(\ref{average}) is 
always positive and can hence be used for importance sampling. This is crucial 
for a complete solution of the complex action problem.

For a given bond configuration one can integrate out the spin variables and one
obtains
\begin{eqnarray}
\label{Zbond}
Z&=&\int {\cal D}b \ (e^\kappa - 1)^{N_b} \ 3^{N_C} \prod_C W(C) 
\nonumber \\
&=&\int {\cal D}b \ (e^\kappa - 1)^{N_b} \prod_C
[\exp(h |C|) + 2 \exp(- h |C|/2) \cos(\sqrt{3} h |C|/2)].
\end{eqnarray}
Here $N_b$ is the number of bonds that are put (i.e.~have $b = 1$). The factor 
$3^{N_C}$ represents the number of allowed spin configurations for a given bond
configuration and the factors $W(C)$ come from the improved estimator. The 
effective action for the bond variables depends only on the sizes $|C|$ of the 
clusters corresponding to a given bond configuration. Note that the factor 
$1/3$ per cluster in eq.(\ref{average}) cancels against the factor $3^{N_C}$. 

Our algorithm directly updates the partition function of eq.(\ref{Zbond}), 
i.e.~it only operates on the bond variables while the spins are already 
integrated out analytically.\footnote{B. Scarlet was first to realize
that the spin variables need not even be simulated.} The bond variables that 
define the clusters are updated with a local algorithm. A bond whose value does
not affect the cluster sizes is put with probability $p = 1 - e^{- \kappa}$. 
This happens when the two sites at its ends belong to the same cluster because 
they are connected indirectly through other bonds. A bond whose value affects 
the cluster sizes is put with a probability that depends on the sizes of the 
corresponding clusters. When the bond is not put ($b = 0$), its endpoints $x$ 
and $y$ belong to two different clusters $C_1$ and $C_2$ of sizes $|C_1|$ and 
$|C_2|$ and the corresponding Boltzmann weight is $3^2 W(C_1) W(C_2)$. On the 
other hand, when the bond is put ($b = 1$), its endpoints belong to the 
combined cluster $C_1 \cup C_2$ of size $|C_1| + |C_2|$. In that case, the 
Boltzmann weight is $3 W(C_1 \cup C_2) (e^\kappa - 1)$. Hence, the bond is put 
with probability
\begin{equation}
q = \frac{W(C_1 \cup C_2)(e^\kappa - 1)}
{3 W(C_1) W(C_2) + W(C_1 \cup C_2)(e^\kappa - 1)}.
\end{equation}

\subsection{Improved Estimators for Physical Quantities}

In order to measure physical observables, it is crucial to construct improved 
estimators for them as well. Here we construct improved estimators for the 
Polyakov loop $\Phi_x$, its charge conjugate $\Phi_x^*$, as well as for the 
correlators $\Phi_x \Phi_y^*$, $\Phi_x \Phi_y$ and $\Phi_x^* \Phi_y^*$. The
expectation values
\begin{equation}
\langle \Phi_x \rangle = \exp(- \beta F_Q), \
\langle \Phi_x^* \rangle = \exp(- \beta F_{\bar Q}),
\end{equation}
determine the free energies $F_Q$ of a quark and $F_{\bar Q}$ of an anti-quark.
The Polyakov loop correlators determine the quark-anti-quark potential 
$V_{Q \bar Q}(x - y)$, the quark-quark potential $V_{Q Q}(x - y)$ and the
anti-quark-anti-quark potential $V_{\bar Q \bar Q}(x - y)$ via
\begin{eqnarray}
\exp(- \beta V_{Q \bar Q}(x - y)) = \langle \Phi_x \Phi_y^* \rangle, 
\nonumber \\
\exp(- \beta V_{Q Q}(x - y)) = \langle \Phi_x \Phi_y \rangle, \nonumber \\
\exp(- \beta V_{\bar Q \bar Q}(x - y)) = \langle \Phi_x^* \Phi_y^* \rangle.
\end{eqnarray}
The improved estimator for the Polyakov loop is given by the sub-ensemble 
average
\begin{equation}
\langle \Phi_x \exp(h \sum_z \Phi_z) \rangle_{3^{N_C}} = 
\frac{1}{3} \sum_{\Phi \in \Z(3)} \Phi \exp(h |C_x| \Phi)
\prod_{C \neq C_x} W(C),
\end{equation}
where $C_x$ is the cluster that contains the point $x$. Hence, we obtain
\begin{equation}
\langle \Phi_x \rangle = \frac{1}{Z(h)} \int {\cal D}b \
\frac{1}{3 W(C_x)} \sum_{\Phi \in \Z(3)} \Phi \exp(h |C_x| \Phi)
(e^\kappa - 1)^{N_b} \ 3^{N_C} \prod_C W(C),
\end{equation}
i.e.~after integrating out the spin variables, the Polyakov loop is represented
by 
\begin{equation}
\Phi_x = \frac{1}{3 W(C_x)} \sum_{\Phi \in \Z(3)} \Phi \exp(h |C_x| \Phi).
\end{equation}
Similarly, the operator representing the charge conjugate Polyakov loop is
\begin{equation}
\Phi_x^* = \frac{1}{3 W(C_x)} \sum_{\Phi \in \Z(3)} \Phi^* \exp(h |C_x| \Phi).
\end{equation}
The improved estimator for the Polyakov loop correlator $\Phi_x \Phi_y^*$ is 
given by the sub-ensemble average
\begin{eqnarray}
\langle \Phi_x \Phi_y^* \exp(h \sum_z \Phi_z) \rangle_{3^{N_C}}
&=&\frac{1}{3} \sum_{\Phi \in \Z(3)} \Phi \exp(h |C_x| \Phi) \
\frac{1}{3} \sum_{\Phi^* \in \Z(3)} \Phi^* \exp(h |C_y| \Phi) \nonumber \\
&\times&\prod_{C \neq C_x,C_y} W(C),
\end{eqnarray}
if the points $x$ and $y$ belong to two different clusters $C_x$ and $C_y$. If
the points $x$ and $y$ belong to the same cluster (i.e.~if $C_x = C_y$), the
improved estimator is simply given by
\begin{equation}
\langle \Phi_x \Phi_y^* \exp(h \sum_z \Phi_z) \rangle_{3^{N_C}} = 
\prod_{C} W(C),
\end{equation}
because then $\Phi_x \Phi_y^* = 1$. Hence, the operator representing the
correlator is
\begin{eqnarray}
\Phi_x \Phi_y^*&=& 
\frac{1}{9 W(C_x) W(C_y)} \sum_{\Phi \in \Z(3)} \Phi \exp(h |C_x| \Phi)
\sum_{\Phi \in \Z(3)} \Phi^* \exp(h |C_y| \Phi), \ 
\mbox{if} \ C_x \neq C_y, \nonumber \\
\Phi_x \Phi_y^*&=&1, \ \mbox{if} \ C_x = C_y.
\end{eqnarray}
Similarly, we have
\begin{eqnarray}
\Phi_x \Phi_y&=&
\frac{1}{9 W(C_x) W(C_y)} \sum_{\Phi \in \Z(3)} \Phi \exp(h |C_x| \Phi)
\sum_{\Phi \in \Z(3)} \Phi \exp(h |C_y| \Phi), \ 
\mbox{if} \ C_x \neq C_y, \nonumber \\
\Phi_x \Phi_y&=&
\frac{1}{3 W(C_x)} \sum_{\Phi \in \Z(3)} \Phi^2 \exp(h |C_x| \Phi), \ 
\mbox{if} \ C_x = C_y, \nonumber \\
\Phi_x^* \Phi_y^*&=& 
\frac{1}{9 W(C_x) W(C_y)} \sum_{\Phi \in \Z(3)} \Phi^* \exp(h |C_x| \Phi)
\sum_{\Phi \in \Z(3)} \Phi^* \exp(h |C_y| \Phi), \ 
\mbox{if} \ C_x \neq C_y, \nonumber \\
\Phi_x^* \Phi_y^*&=&
\frac{1}{3 W(C_x)} \sum_{\Phi \in \Z(3)} (\Phi^*)^2 \exp(h |C_x| \Phi), \ 
\mbox{if} \ C_x = C_y.
\end{eqnarray}

\subsection{Severity of the Complex Action Problem}

\begin{figure}
\begin{center}
\rotatebox{90}{\hspace{3.7cm}$\langle \exp(i h \sum_x \mbox{Im}\Phi_x) \rangle_R$}
\epsfig{file=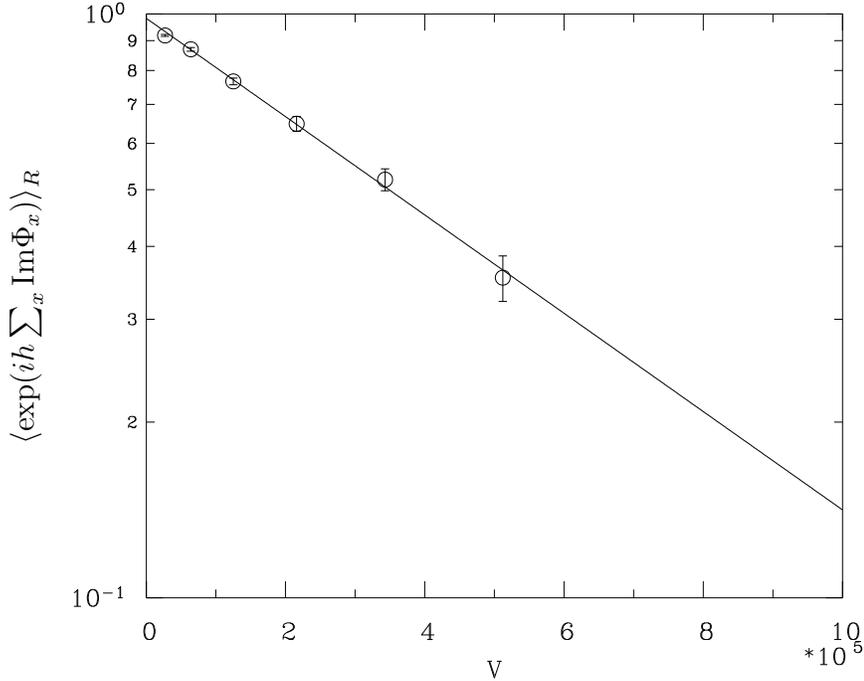,
width=9.5cm,angle=90,
bbllx=50,bblly=140,bburx=535,bbury=740}
\end{center} 
\vspace*{-0.6cm}
\caption{\it The expectation value of the phase factor as a function
of the volume at the critical endpoint $E$.}
\end{figure}
In order to estimate the severity of the complex action problem, we like to 
determine the expectation value of the complex phase of the Boltzmann factor in
the modified real action ensemble
\begin{equation}
\langle \exp(i h \sum_x \mbox{Im}\Phi_x) \rangle_R = \frac{Z}{Z_R}.
\end{equation}
Rather than implementing this directly in a simulation that uses the absolute 
value of the Boltzmann factor for importance sampling, one can measure $Z_R/Z$ 
with the cluster algorithm. In fact, an improved estimator for this quantity is
given by $\prod_C W_R(C)/W(C)$, where
\begin{equation}
W_R(C) = \langle \exp(h \sum_{x \in C} \mbox{Re}\Phi_x) \rangle_3 =
\frac{1}{3}[\exp(h |C|) + 2 \exp(- h |C|/2)],
\end{equation}
is the weight that replaces $W(C)$ in the real action ensemble. Alternatively,
one can construct a cluster algorithm that simulates the real action ensemble.
In that case, one needs to measure $\prod_C W(C)/W_R(C)$ in order to obtain 
$Z/Z_R$. 

\begin{figure}
\begin{center}
\epsfig{file=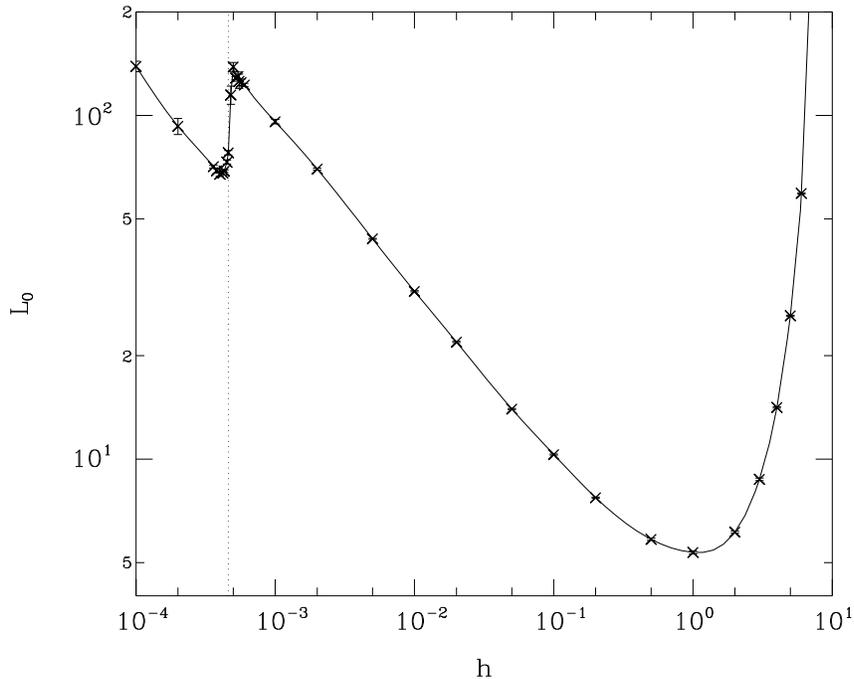,
width=9.5cm,angle=90,
bbllx=50,bblly=200,bburx=535,bbury=800}
\end{center} 
\vspace*{-0.6cm}
\caption{\it The scale parameter $L_0$ related to the severity of the complex action
problem as a function of $h$ at $\kappa=0.5495$. The solid line is a spline
to guide the eye. The phase transition takes place at the dotted
vertical line.}
\end{figure}
Figure 3 shows $\langle \exp(i h \sum_x \mbox{Im}\Phi_x) \rangle_R$ as
a function of the volume $V = L^3$ at the critical endpoint of the transition
line (point $E$ in figure 1). Indeed, one finds an exponentially small signal, as
expected from eq.(\ref{phase}). Defining a scale parameter $L_0$
that measures the severity of the complex action problem by
\begin{equation}
\langle \exp(i h \sum_x \mbox{Im}\Phi_x) \rangle_R\propto\exp\left(-\frac{L^3}{L_0^3}\right)
\end{equation}
we get $L_0\approx 80$ at the endpoint $E$. This means that the
complex action
problem at the endpoint is extremely mild. In fact, in practice it 
is not a problem at all up to volumes as big as $100^3$. Since the
current computer hardware restricts the system size to a couple of
million degrees of freedom, one can also study the point $E$ 
with an algorithm that does not solve the complex action problem. On very
large lattices the meron-cluster algorithm will become superior also
at the point $E$ since
its computational effort is polynomial in the system size as opposed to
exponential for the reweighted Metropolis algorithm.

It should be noted that the complex action problem is most severe for 
intermediate values of $h$. While it is obvious that there is no complex action
problem at $h = 0$, it is perhaps less obvious that there is also no problem 
for large $h$. This is because
\begin{equation}
\frac{W_R(C)}{W(C)} = \frac{\exp(h |C|) + 2 \exp(- h |C|/2)}
{\exp(h |C|) + 2 \exp(- h |C|/2) \cos(\sqrt{3} h |C|/2)}
\end{equation}
approaches $1$ in the limit $h \rightarrow \infty$ so that 
$\langle \exp(i h \sum_x \mbox{Im}\Phi_x)_R \rangle_R = Z_R/Z \rightarrow 1$.
Figure 4 shows the complex action problem scale parameter $L_0$ as a function of
$h$ for fixed $\kappa=0.5495$. It has a minimum at $h\approx 1$
meaning that the complex action problem ist most severe in that
region.
Defining the ``practical complex action problem'' as being present when 
$\langle \exp(i h \sum_x \mbox{Im}\Phi_x) \rangle_R<0.01$ on a $100^3$
lattice (i.e.~$L_0<60$) we see that at $\kappa=0.5495$ there is no
practical complex action problem for $h<0.003$ as well as for $h>6$.
A more physical definition of a practical complex action problem would
compare $L_0$ with the correlation length $\xi$ but we have not
measured the correlation length.

\section{Flux Representation of the Potts Model}

In this section we map the Potts model to an equivalent flux model that does 
not suffer from the complex action problem. Then we describe a Metropolis
algorithm to update the flux model and compare its efficiency with the cluster
algorithm.

\subsection{Mapping the Potts Model to a Flux Model}

As pointed out in \cite{Con00} the Potts model can be rewritten as a flux model
that does not suffer from the complex action problem. As we have seen, the 
complex action problem can also be solved in the cluster formulation. Hence,
the question arises if the flux or the cluster formulation leads to more 
efficient numerical simulations. Let us first match the flux model to the 
original Potts model. The flux model is formulated in terms of ``electric'' 
charges $Q_x \in \{0,\pm 1\}$ defined on the lattice sites $x$ and electric 
flux variables $E_{x,i} \in \{0,\pm 1\}$ living on the links. The charge and 
flux variables are related by the $\Z(3)$ Gauss law constraint
\begin{equation}
\label{Gauss}
Q_x = \sum_i (E_{x,i} - E_{x - \hat i,i}) \mbox{mod} \ 3.
\end{equation}
The action of the flux model takes the form
\begin{equation}
\label{action}
S[E,Q] = \frac{g^2 }{2}\sum_{x,i} E_{x,i}^2 + \beta \sum_x (M Q_x^2 - \mu Q_x).
\end{equation}
The mass $M$ and the chemical potential $\mu$ of the $\Z(3)$ charges are not
directly related to the mass and chemical potential of quarks in QCD (which we 
also denoted by $M$ and $\mu$ in section 2) but qualitatively they play the 
same role. The partition function of the flux model takes the form
\begin{equation}
\label{partition}
Z = \prod_x \sum_{Q_x \in \{0,\pm 1\}} 
\prod_{x,i} \sum_{E_{x,i} \in \{0,\pm 1\}} \prod_x \delta_x \exp(- S[E,Q]).
\end{equation}
The $\delta$-function $\delta_x$ imposes the Gauss law of eq.(\ref{Gauss}) at 
the point $x$ and can be written as
\begin{equation}
\delta_x = \frac{1}{3} \sum_{\Phi_x \in \Z(3)} 
\Phi_x^{Q_x - \sum_i (E_{x,i} - E_{x - \hat i,i})}.
\end{equation}
Inserting this as well as eq.(\ref{action}) for the action in 
eq.(\ref{partition}) one can integrate out the $E_{x,i}$ and $Q_x$ variables. 
The result of the $E_{x,i}$ integration is
\begin{equation}
\sum_{E_{x,i} \in \{0,\pm 1\}} (\Phi_x^* \Phi_{x + \hat i})^{E_{x,i}} 
\exp\left(- \frac{g^2}{2} E_{x,i}^2\right) =
1 + 2 \exp\left(- \frac{g^2}{2}\right) \mbox{Re}(\Phi_x^* \Phi_{x + \hat i}).
\end{equation}
In the Potts model (up to an overall factor) the corresponding term is
$\exp(\kappa \delta_{\Phi_x,\Phi_{x + \hat i}})$. Thus, the flux model matches
the Potts model if
\begin{equation}
\exp(\kappa) = \frac{1 + 2 \exp(- g^2/2)}{1 - \exp(- g^2/2)}.
\end{equation}
Note that the $g \rightarrow 0$ limit of the flux model corresponds to the
$\kappa \rightarrow \infty$ limit of the Potts model. When one integrates out 
the charges $Q_x$ one obtains
\begin{equation}
\sum_{Q_x \in \{0,\pm 1\}} \Phi_x^{Q_x} \exp(- M Q_x^2 - \mu Q_x) =
1 + \exp(- \beta(M - \mu)) \Phi_x + \exp(- \beta(M + \mu)) \Phi_x^*.
\end{equation}
In the original Potts model (up to an overall factor $A$) the corresponding 
term is $\exp(h \Phi_x)$. Hence, the flux model matches the Potts model if
\begin{eqnarray}
&&\frac{1}{3} \sum_{\Phi \in \Z(3)} \exp(h \Phi) = A, \nonumber \\
&&\frac{1}{3} \sum_{\Phi \in \Z(3)} \Phi \exp(h \Phi) = 
A \exp(- \beta(M + \mu)), \nonumber \\
&&\frac{1}{3} \sum_{\Phi \in \Z(3)} \Phi^* \exp(h \Phi) = 
A \exp(- \beta(M - \mu)).
\end{eqnarray}
These relations can be used to determine the parameters 
$\exp(- \beta(M - \mu))$ and $\exp(- \beta(M + \mu))$ of the flux model in 
terms of the Potts model parameter $h$.

\subsection{Metropolis Algorithm for the Flux Model and Comparison with the
Cluster Algorithm}

As first described in \cite{Con00}, the flux model can be updated with a 
simple Metropolis algorithm. One basic move in the algorithm creates or
annihilates a nearest-neighbor charge-anti-charge pair across a given link. The
other basic move creates or annihilates an electric flux loop around an
elementary plaquette. These basic moves are proposed on every link and 
plaquette and are accepted or rejected in a Metropolis step. We have 
implemented both the Metropolis algorithm for the flux model and the cluster 
algorithm for the Potts model and we have verified that physical observables 
obtained with the two algorithms agree with each other. 

The question arises which of the two algorithms is more efficient. The 
Metropolis algorithm is expected to suffer from critical slowing down at the 
endpoint of the first order phase transition with a dynamical critical exponent
$z \approx 2$. Cluster algorithms are known to drastically reduce critical
slowing down, in some cases even to $z \approx 0$. However, our cluster 
algorithm cannot eliminate critical slowing down completely because the 
decision to put a bond is more time-consuming than the one in the Swendsen-Wang
algorithm. For example, when a bond is removed, one must check if an old 
cluster decomposes into two new clusters. To minimize the computational effort,
we simultaneously grow two clusters from the two ends of the bond. Once they 
touch each other, we know that the old cluster did not decay. Still, in the
less likely event that the old cluster does decay, one must completely grow the
smaller of the two clusters in order to decide if the bond can be deleted. Here
we do not attempt to determine the dynamical critical exponent $z$ of our 
cluster algorithm at the critical endpoint.

\begin{figure}
\vspace{-3.0cm}
\begin{center}
(a)\epsfig{file=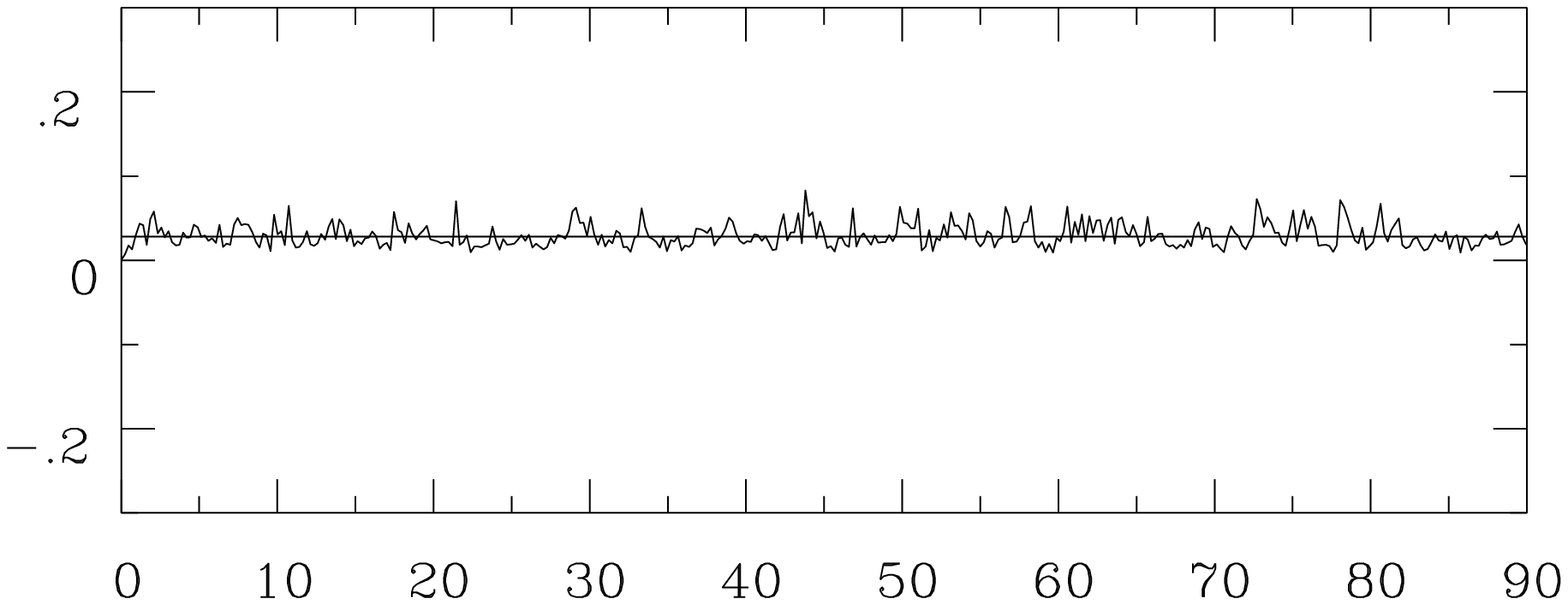,
width=5.0cm,angle=90,
bbllx=210,bblly=190,bburx=420,bbury=720}
(b)\epsfig{file=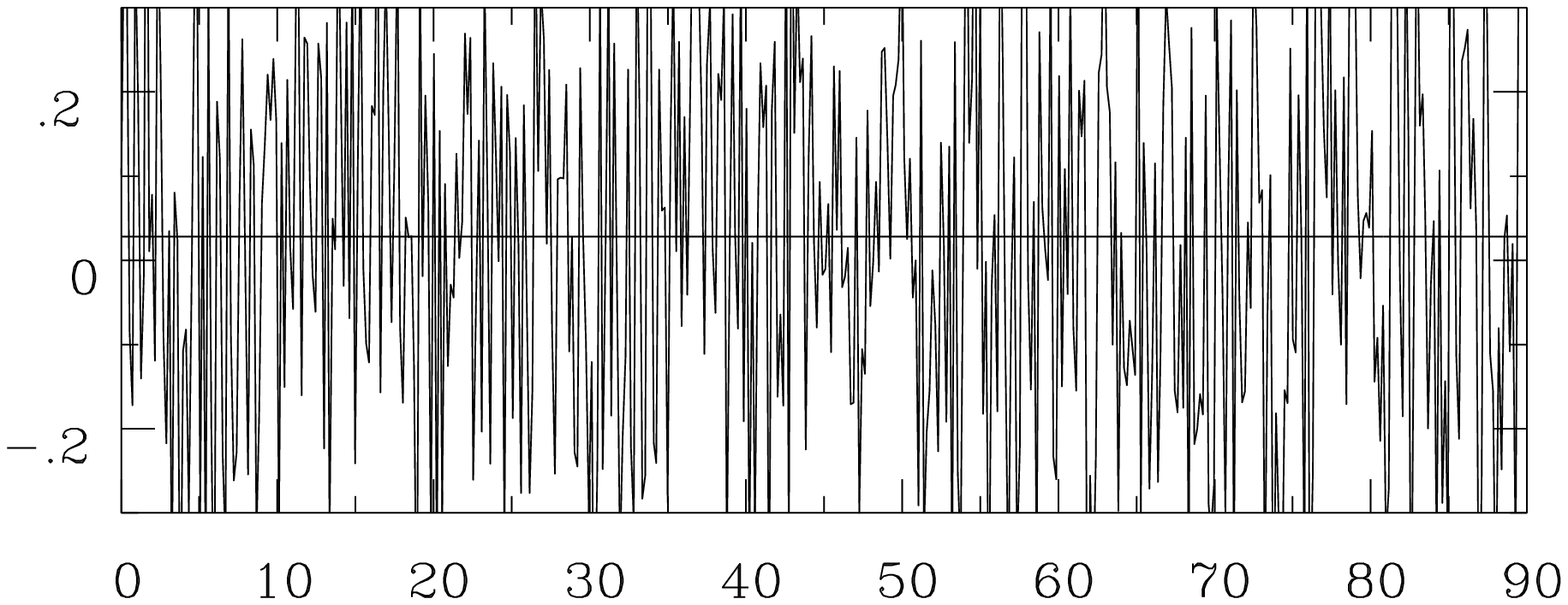,
width=5.0cm,angle=90,
bbllx=210,bblly=190,bburx=420,bbury=720}
(c)\epsfig{file=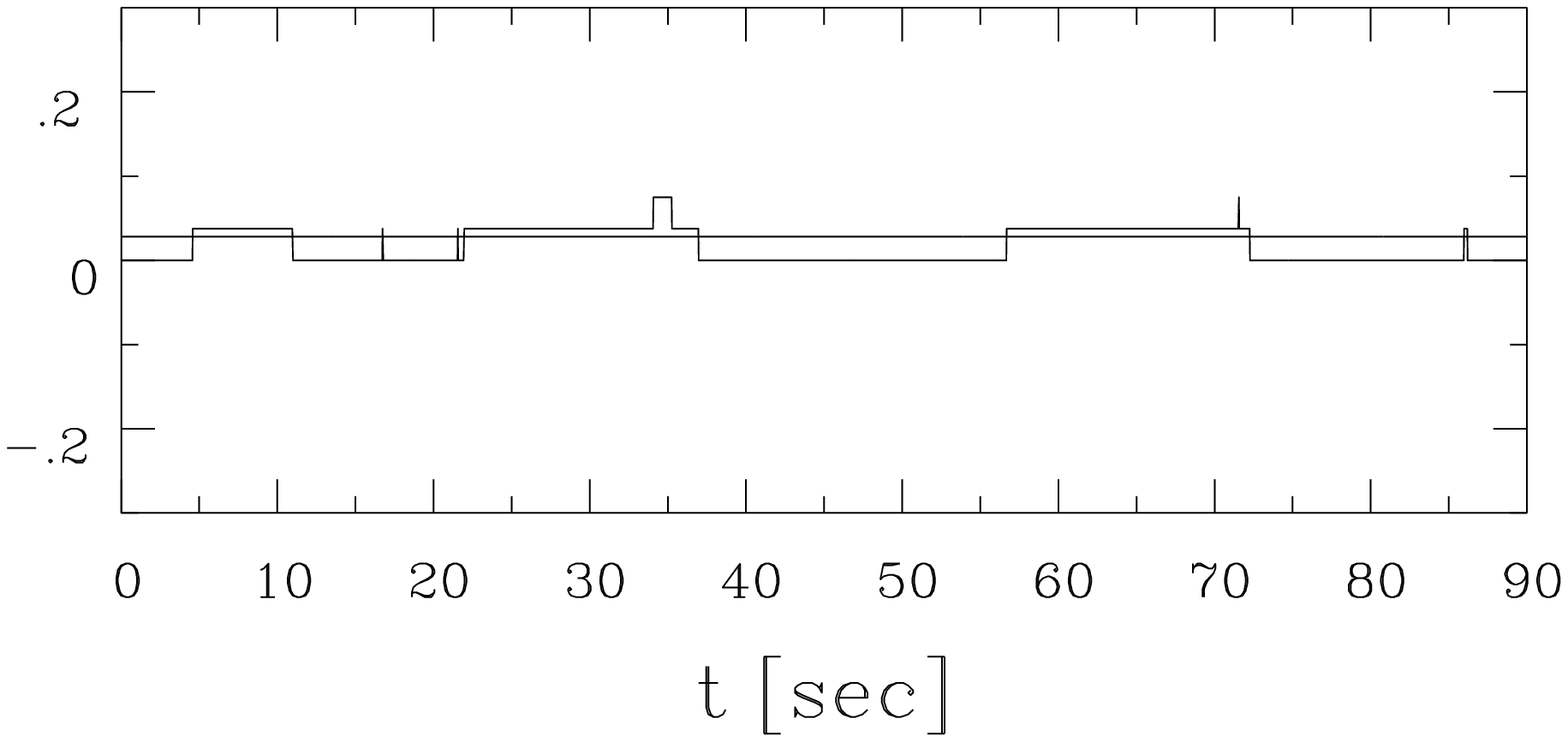,
width=5.0cm,angle=90,
bbllx=210,bblly=190,bburx=420,bbury=720}
\end{center} 
\vspace*{4.0cm}
\caption{\it The computer time history of the Polyakov loop $\Phi$ (a) for the
cluster algorithm for the Potts model using the improved estimator for
$\Phi$, (b) for the Metropolis
algorithm for the Potts model using $\Phi$ directly, 
and (c) for the Metropolis algorithm for 
the flux model on a $20^3$ lattice at $(h,\kappa)=(0.01,0.5)$. 
The horizontal straight line denotes the expectation value.
The Metropolis algorithm for the flux model has a much longer
autocorrelation time than the other two algorithms.
In the case of the Metropolis algorithm for the Potts model the complex
action problem manifests itself by large fluctuations around the
expectation value. 
}
\end{figure}

An even more severe super-critical slowing down is expected close to the first 
order phase transition line. Along that line, the deconfined phase coexists
with the confined phase, and the Monte Carlo simulation must tunnel between the
two phases. In order to tunnel between the confined and the deconfined phase, a
local algorithm must go through configurations containing both phases 
simultaneously. Since the interface that separates the two phases has non-zero
interface tension, such configurations are exponentially suppressed. Hence, 
close to a first order phase transition, a local algorithm like the Metropolis 
algorithm necessarily suffers from exponential slowing down. This 
super-critical slowing down is even more severe than the power-law critical 
slowing down at a second order phase transition. Hence, we expect that the 
Metropolis algorithm for the flux model is not very well suited to study the 
first order phase transition line. In some cases, cluster algorithms can even
eliminate super-critical slowing down. For example, in the broken phase of the
Potts model three distinct deconfined phases coexist with each other. The
Swendsen-Wang algorithm can efficiently tunnel from one deconfined phase to 
another because it assigns the same random $\Z(3)$ element to all spins in a
cluster in a non-local spin update. Still, even the Swendsen-Wang algorithm
suffers from super-critical slowing down at the first order phase transition
that separates the confined from the deconfined phase. Although cluster flips
can naturally lead to tunneling between distinct deconfined phases, they do not
lead directly from a deconfined to the confined phase. To cure this problem,
Rummukainen \cite{Rum93} has combined the cluster algorithm with the
multi-canonical methods of Berg and Neuhaus \cite{Ber91} which can reduce the
exponential super-critical slowing down to a power-law behavior. Although this 
may well be possible, we have not yet attempted to combine our algorithm with 
multi-canonical methods. Hence, we expect that our cluster algorithm still 
suffers from super-critical slowing down close to the deconfinement phase 
transition. 

We have compared the efficiency of the Metropolis algorithm for the flux model,
the cluster algorithm for the complex action Potts model, and the
reweighted Metro\-po\-lis algorithm for the complex action Potts model at
several points in the phase diagram.
In figure 5 we compare the computer time 
histories of the Polyakov loop for the three algorithms mentioned
above  at $(h,\kappa)=(0.01,0.5)$ which is in the confined region of
the phase diagram.
Obviously, the flux model Metropolis algorithm decorrelates a lot
worse than the other two algorithms.
 The flux algorithm performs even worse when one approaches the first
order transition line. The reweighted Metropolis algorithm for the
Potts model suffers from the complex action problem. What is
plotted is the time evolution of 
${\rm Re}[\sum_x\Phi_x \exp(i h \sum_x \mbox{Im}\Phi_x)]/V 
\langle \exp(i h \sum_x \mbox{Im}\Phi_x) \rangle_R$. 
Its statistical fluctuations are much larger than the
expectation value which is a manifestation of the complex action problem.

\section{Universality Class of the Critical Endpoint}

In this section we present the results of our numerical simulations at
the critical endpoint. The complex action problem turned out to be very weak in
the vicinity of the critical endpoint (see subsection 3.4). Therefore 
it is possible to use a simple reweighted Metropolis algorithm
even though that does not solve the complex action problem. It is usable up to
sufficiently large lattices so that critical exponents can be
extracted from a finite size scaling analysis. Of course, on even larger
lattices the meron-cluster algorithm will eventually be
superior to the reweighted Metropolis algorithm. But since at the
endpoint $E$ the complex action problem sets in only at volumes $\gtrsim 100^3$, 
simulations at $E$ are not limited by the complex action problem but by the
ability to simulate large lattices on today's computers. 

Figure 1 shows the phase diagram of the model defined by equation (\ref{Potts}). For 
$h = 0$ our model reduces to the standard 3-d 3-state Potts model which has 
been studied extensively in Monte Carlo simulations \cite{Jan97}. The model is 
known to have a weak first order phase transition. The value of the coupling 
$\kappa$ where the phase transition occurs (point $T$ in fig.~1) has been 
determined with high precision. In \cite{Jan97}, the phase transition was 
found to occur at $\kappa_T = 0.550565(10)$. Above this value the $\Z(3)$ 
symmetry is spontaneously broken, i.e.~for $\kappa > \kappa_T$ three distinct
deconfined phases coexist. When we switch on the parameter $h$, the $\Z(3)$ 
symmetry gets explicitly broken. Positive values of $h$ favor the deconfined
phase with a real value of $\langle \Phi \rangle$. Hence, the line $\kappa >
\kappa_T$ at $h = 0$ is a line of first order phase transitions which cannot
terminate in the deconfinement transition at the point $T$. In fact, $T$ is a
triple point because two other first order transition lines emerge from it. For
$h > 0$ a line of first order transitions extends into the $(h,\kappa)$-plane
and terminates in a critical endpoint ($E$ in fig.~1). Negative values of $h$
favor the two deconfined phases with complex values of $\langle \Phi \rangle$. 
Negative $h$ are unphysical in the QCD interpretation of the Potts model 
because $h$ represents $\exp(- \beta(M - \mu))$ in QCD. Still, the Potts model 
at $h < 0$ makes perfect sense as a statistical mechanics system (unrelated to 
QCD) and it has another first order transition line emerging from the point 
$T$. Interestingly, with our method the complex action problem can only be solved for 
$h \ge 0$ since otherwise the improved estimator of eq.(\ref{average}) is not 
necessarily positive. It should be noted that for $h < 0$ also the flux model
suffers from the complex action problem.

The line of first order phase transitions $\kappa_t(h)$ is determined by the
condition that the free energy densities of the confined and deconfined phases
are equal, i.e.~$f_c(h,\kappa_t(h)) = f_d(h,\kappa_t(h))$. Close to
the point $T = (0,\kappa_T)$ the free energy density of the confined phase is 
given by
\begin{equation}
f_c(h,\kappa) = f_{c,T} + e_{c,T}(\kappa - \kappa_T),
\end{equation}
where $f_{c,T} = f_c(0,\kappa_T)$ and $e_{c,T} = df_c/d\kappa(0,\kappa_T)$ is 
the energy density of the confined phase at the point $T$. Note that to leading
order $f_c(h,\kappa)$ is independent of $h$ because $\langle \Phi \rangle = 0$ 
in the confined phase at $h = 0$. On the other hand, for the deconfined phase 
one obtains
\begin{equation}
f_d(h,\kappa) = f_{d,T} + e_{d,T}(\kappa - \kappa_T) - 
h \langle \Phi \rangle_T,
\end{equation}
where $\langle \Phi \rangle_T$ is the value of the Polyakov loop at the point
$T$ in the deconfined phase that is favored at $h > 0$. Using the condition 
$f_{c,T} = f_{d,T}$ for the deconfinement phase transition at $h = 0$, one 
finds
\begin{equation}
\label{line}
\kappa_t(h) =\kappa_T- \frac{\langle \Phi \rangle_T}{e_{c,T} - e_{d,T}} 
h =\kappa_T- \frac{h}{r}.
\end{equation}
The Monte Carlo data of \cite{Jan97} and \cite{GaKa89}
imply $r =0.41(1).$
Our data are
consistent with the first order transition line being a straight line. 
Fitting the values of $\kappa_t(h)$ obtained from the infinite volume 
extrapolation described below yields $r = 0.430(6)$
in reasonable agreement with the number from above. 
Similarly, one can determine the
angle at which the third transition line leaves the point $T$ in the direction
of negative $h$. A similar argument can be applied to the Potts model with real
action that was studied in \cite{Kar00}. Also in that case the first order 
phase transition line is consistent with a straight line and again the 
predicted position for the transition line agrees with the numerical data. 
Interestingly, for the Potts model with both real and complex action, 
information at $h = 0$ is sufficient to predict the position of the transition 
line for $h > 0$. This is because the transition at $h = 0$ is rather weak and 
the line ends already at small values of $h$. If the transition would extend 
deep into the $(h,\kappa)$-plane one would expect deviations from a straight 
line that would be hard to predict based on data at $h = 0$. 

\begin{figure}
\begin{center}
\epsfig{file=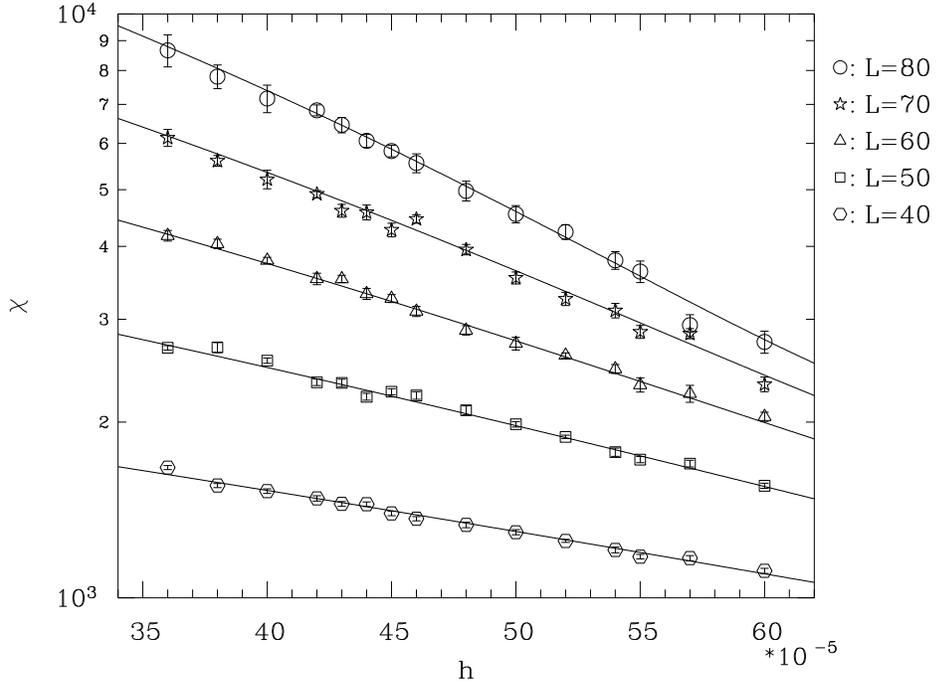,
width=9.5cm,angle=90,
bbllx=50,bblly=200,bburx=535,bbury=800}
\end{center} 
\vspace*{-0.6cm}
\caption{\it The susceptibility $\chi$ along the transition line
plotted as a function of $h$ for five different volumes.}
\end{figure}

To determine the location of the transition line numerically we
perform for given values of $h$ and the volume $V$ simulations at 3 to
5 different values of $\kappa$. These simulations are then combined with
Ferrenberg Swendsen multi-histogram reweighting \cite{FeSw}. To estimate the position of the
transition line we use the specific heat
\begin{equation}
C=\frac{1}{V}\left(\langle (S[\Phi]-h\sum_x\Phi_x )^2 \rangle
-
\langle (S[\Phi]-h\sum_x\Phi_x ) \rangle^2\right)
\end{equation}
and determine the position of its maximum $\kappa_t(h,V)$ 
for a given $h$ and $V$.
The transition point $\kappa_t(h)$ 
is determined in the infinite volume limit using
\begin{equation}
\kappa_t(h,V)=\kappa_t(h)+\frac{A(h)}{V}
\end{equation}
where $\kappa_t(h)=\kappa_t(h,V=\infty)$. This ansatz is used
sucessfully in the whole $h$-range, i.e.~for the first order 
region as well as in the crossover region. The values for $\kappa_t(h)$ 
are plotted with errorbars into the phase diagram (figure 1). On the 
first order transition line they are consistent with a straight 
line which intersects the $\kappa$ axis exactly at the point $T$.
The crossover line has a slight curvature.

After we have determined the tangent to the first order transition
line close to the endpoint we still have to find the exact location of
the endpoint on that line. Also we want to extract critical
exponents. For that purpose we consider the Polyakov loop
susceptibility  
\begin{equation}
\chi=\frac{1}{V}\left(\langle (\sum_x\Phi_x )^2 \rangle
-
\langle \sum_x\Phi_x  \rangle^2\right)
\end{equation}
along the transition line, i.e. at the points
$(\kappa_t(h),h)$. One could consider a ``rotated susceptibility''
instead, where an admixture of the kinetic energy term is added to the
Polyakov loop to diagonalize the fluctuation matrix. Nevertheless, this
is not necessary, since the ``magnetic field direction'' is the
dominant one. We explicitly checked that the results do not depend on
the admixture we chose, unless one comes close to the linear
combination where the discontinuities of the Polyakov loop and the
kinetic energy cancel. This linear combination
would correspond to the kinetic energy term in the Ising model.
Of course, if one would be interested in observables related to the
kinetic energy of the Ising model, one would have to exactly choose
the linear combination that corresponds to the kinetic energy direction.

Close to the critical point the following scaling ansatz describes the 
susceptibility (see e.g.~\cite{Bl95})
\begin{equation}
\chi=L^{\gamma/\nu}f(x),\quad x=(h-h_c)L^{1/\nu}.
\end{equation}
For the fit, the function $f(x)$ is expanded in a Taylor series around
$x=0$ up to second order, $f(x)=f_0+f_1x+f_2x^2/2$. 
We perform two fits:
\begin{itemize}
\item Six parameter fit: It results in 
$1/\nu=1.532(57)$, 
$\gamma/\nu=2.064(70)$, 
$h_c=0.000445(18)$, 
$f_0=0.70(16)$, 
$f_1=-3.99(95)$, and
$f_2=16.4(60)$ with $\chi^2/{\rm d.o.f.}=1.33$. The results for the exponents
agree with the
estimates for the 3d-Ising universality class  $1/\nu=1.587(2)$
and $\gamma/\nu=1.963(3)$ \cite{Bl95} almost within errorbars.

\item Four parameter fit: The critical exponents are fixed
to the Ising model values.
The result is $h_c=0.000470(2)$, $f_0=0.9775(46)$, $f_1=-4.567(55)$, and
$f_2=16.5(17)$ with $\chi^2/{\rm d.o.f.}=1.35$. The good value for $\chi^2$
supports again the universality class of the 3-d Ising model.
\end{itemize}
\begin{figure}
\begin{center}
\epsfig{file=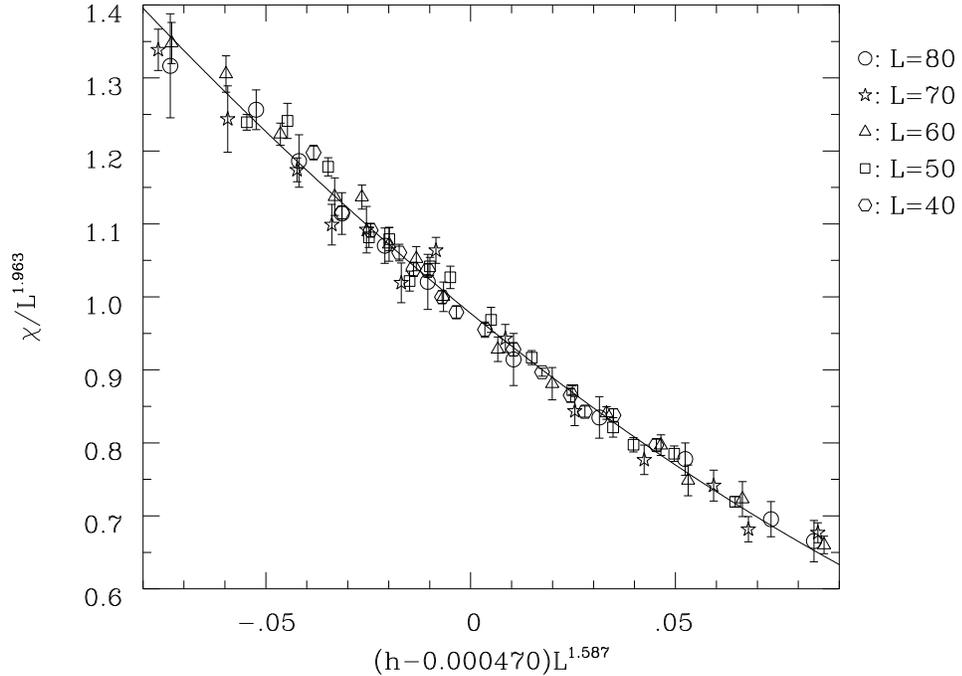,
width=9.5cm,angle=90,
bbllx=50,bblly=200,bburx=535,bbury=800}
\end{center} 
\vspace*{-0.6cm}
\caption{\it The scaling function $f(x)$.}
\end{figure}
The susceptibility and the four parameter fit are shown in Figure 6.
Figure 7 shows the function $f(x)$.

\section{Conclusions}

We have used a cluster algorithm to solve the notorious complex action
problem in the Potts model approximation to
QCD with heavy quarks at large chemical
potential. We use a simple analytically constructed improved
estimator that gives an exponential reduction in the required
statistics. Since the improved estimator is real and positive,
importance sampling techniques that fail for complex actions then
become applicable. This makes it possible to study the 
whole $h>0$ parameter range of the Potts model, not just the
$h=0$ axis. 
(Recall that $h$ corresponds to $\exp(\beta(\mu-M))$ 
in QCD in the limit $M,\mu \to \infty$ at any given $\mu - M$).

We compared our cluster algorithm with a flux model reformulation,
and with the reweighted Monte Carlo algorithm.
We found that the cluster algorithm was more efficient than using
the flux model reformulation.
In the large volume limit the cluster algorithm will always be superior
to Monte Carlo reweighting. 
However, at very small $h$ 
it is sufficient to use reweighting techniques to obtain physically
 relevant results.
This turned out to be the case for the endpoint of the
first order line, which occurs at a very small $h$ because the
3-d 3-state Potts model phase transition is rather weak at $h=0$.
We therefore used reweighted Monte Carlo to locate the 
first-order line and its endpoint.
However, we emphasize that as computer power rises, and the
maximum attainable volume becomes bigger, the meron-cluster
algorithm will eventually become superior at any $h>0$.

We also calculated quark-quark, quark-antiquark, and antiquark-antiquark
potentials, in the confined and deconfined regions of the phase diagram.
We found the expected behavior: the background density of
heavy quarks screens color fields, so that all potentials reach plateaux
at long distances, whose values are simply related to the free
energies of external static quarks and antiquarks.

The algorithm that we have developed for the Potts model belongs to
the class of meron-cluster algorithms that has recently been used to
solve a large variety of sign and complex action problems. Of course,
the ultimate goal is to construct a similar algorithm for QCD at
non-zero chemical potential and investigate the phase structure of QCD
at $\mu \neq 0$ from first principles.  The complex action problem in
full QCD is more complicated than the one in the Potts model.
So far, meron-cluster
algorithms have led to solutions of fermion sign problems as well as
complex action problems in bosonic theories, 
but have not yet solved complex action problems
in theories with fermions.
We believe that this may ultimately become possible when one
uses the D-theory formulation of QCD.

\section*{Acknowledgements}

We like to thank the INT in Seattle, where this work was initiated, for its 
hospitality. U.-J.~W.~thanks F.~Karsch and S.~Stickan for helpful discussions
and acknowledges the support of the A.~P.~Sloan 
foundation.

\end{document}